\documentclass[aps,preprint,showpacs,superscriptaddress,groupedaddress]{revtex4}  % for double-spaced preprint
\usepackage{graphicx}  % needed for figures
\usepackage{dcolumn}   % needed for some tables
\usepackage{bm}        % for math
\usepackage{amssymb}   % for math
\begin{document}

\title{Electron energization  in upstream of  collisionless electron/ion shocks produced by interpenetrating plasmas}
\author{N. Naseri$^{1,2}$, V. Khudik$^{1,3}$ and G. Shvets$^{1,3}$\\ %, V. Khudik$^1$ and G. Shvets$^3$
\textit{$^1$ Institute for Fusion Studies and Department of Physics, The University of  Texas, 1 University Station C1500,
Austin, Texas 78712, USA. \\$^2$ Physics and Astronomy Department, Middle Tennessee State University, Wiser-Patten Science Hall, 422 Old Main Cir, Murfreesboro, TN 37132, U.S.A\\
$^3$ School of Applied and Engineering Physics, Cornell University, Ithaca, New York, 14850, USA.} }
\begin{abstract}

We discuss the mechanism of electron energization in the upstream region of relativistic e/i shock. By using particle-in-cell simulations, we demonstrate the electrons interacting with enhanced electric and magnetic fields of the magnetic vortices (MVs)  can gain a significant amount of energy during interaction. MVs are self generated in the upstream region of relativistic electron-ion shock. 

\end{abstract}


\maketitle
\section{Introduction} \label{sec:intro}
Particle acceleration is one of the fundamental topics in astrophysical and laboratory shocks. Collisionless shocks are considered responsible for plasma energization mechanisms leading to relativistic particles. The spectrum of the radiation emitted by high energy particles from indirect observations and by measurements of Cosmic Ray (CR)  and gamma-ray bursts spectrums  show the evidences of non-thermal particle acceleration generated by collisionless shocks[\cite{Piran}]%), Gehrels}).
  These observations  as well as numerical simulations of unmagnetized relativistic collisionless shocks have shown electron heating and energization in upstream region of the electron/ion shock \cite{Panait,Piran,Gehrels,Spitkovsky2008}. 
Indirect observations and numerical simulations  also indicate that the order of magnetic field at the shock front is substantially larger than the intersellar magnetic field, suggesting the instabilities generated by streaming plasmas as a source of amplification of magnetic field. A first phase of amplification happens in the shock front formation. A successive stage of magnetic field growth happens due to secondary streaming instabilities, in particular by the development of Weibel instability driven by particles moving ahead of shock front (counter streams) and the the incoming plasma streams. The investigation on magnetic field growth is still an active research area since these magnetic islands are the source of  particle acceleration and particle transport across universe. Weibel instability in Weibel-mediated collisionless shocks leads to plasma isotropization and to particle energization at later times.
Recently we demonstrated that magnetic vortices (MVs) can self-consistently emerge as a result of the collisionless interaction of two inter-penetrating relativistic plasma streams (electrons and ions) with no external magnetic field \cite{naseri2018}.  Localized regions of the strong magnetic field in the form of magnetic dipole vortices upstream of the shock are observed in the simulation developed during the nonlinear evolution of the electron and ion filaments. However, the interaction of magnetic vortices with particles and consequent particle acceleration requires detailed investigation of vortices evolution and particle dynamics.  Large scale simulations such as Ref.\cite{Spitkovsky2008} revealed the importance of upstream electron energization in establishing the shock transition region. The electron energy spectrum of slices in the upstream of the shock showed a power law spectrum shown in Ref. \cite{Spitkovsky2008}. Although electron energization in shock transition\cite{Spitkovsky2008} region has been extensively studied, the mechanism of electron heating and energization in upstream of the shock is still unclear.
 In this letter we focus on  electron heating and energization mechanism in the foreshock region using 2D particle-in-cell simulation results.  
 %To our knowledge this is the first detailed study of particle energization in the upstream of electron-ion shock.
  We show that the incoming and counter stream electron flows gain significant amount of energy while interacting with nonlinear stage of self-generated magnetic dipole vortices in the upstream region. Tracking a large number of electrons from the tail of the energy spectrum shows that $90\%$ of the energized electrons from the non-thermal tail of the energy spectrum move towards (or return) to shock transition region with energies more than an order of magnitude larger than their initial energy. An estimate for the maximum electron energy is given.
  \begin{figure}
\begin{center}
%\hspace*{-0.4in}
\includegraphics[width=\linewidth]{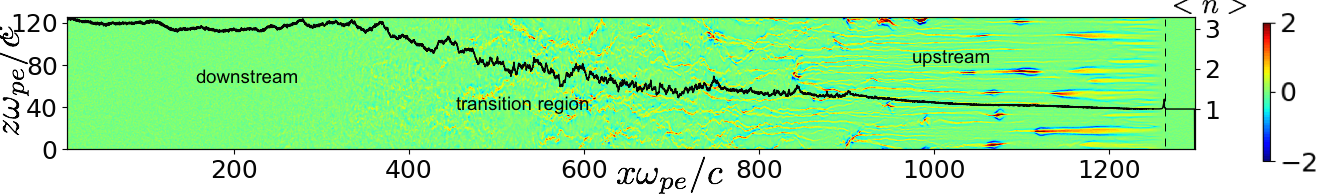}%{sketch4.png}%{magnetic-d.png}
\end{center}
\caption{Schematic of the interaction of two interpenetrating electron/ion beams: Color plot of normalized out of plane magnetic field ($B_y$)  in $x-z$ plane at $\omega_{pe}t=1140$. Solid black curve (axis on right): transversely averaged density $\tilde{N}$ normalized to upstream unperturbed density $n_0$. The dashed line is where the incoming e/i beams meet the counter streams.}\label{fig0}
\end{figure}
%%%%%%%%%%%%%%%%%%%%%%%%%
\section{PIC simulation set-up}
The two-dimensional (2D) version of the PIC fully relativistic parallel simulation code VLPL is used \cite{Pukhov1999}. The code was  modified to minimize noise properties of numerical instabilities, by using third-order shaped particles and current smoothing.
A rectangular simulation box in the $x-z$ plane with the dimensions $L_x = 1300~l_{pe}$ and  $L_z = 130~l_{pe}$ and the grid sizes $\Delta z = l_{pe}/10$ and $\Delta x =  l_{pe}/10$ is used. Here $l_{pe}=c/\omega_{pe}$ is  the electron inertial length, that is the typical transverse spacial scale of the filaments,  $\omega_{pe}=\sqrt{\frac{4\pi n_0e^2}{\gamma m_e}}$ is the electron plasma frequency, $\gamma$ is the relativistic gamma factor of incoming plasma flow, $e$ and $m$ denote the charge and mass of electron, and $n_0$ is the unperturbed density of electrons. Periodic boundary conditions are applied for particles and fields in the transverse ($z$) direction. Each computational cell is initialized with $16$ macro-particles: $8$ electrons and $8$ ions. We assume that initially the electron-ion (e,i) plasma beam with the mass ratio $m_i/m_e=32$ and equal charges $q_i=q_e  $, equal densities $n_0$, and relativistic velocities $v_x$ (corresponding to $\gamma \equiv 1/\sqrt{1-v_x^2/c^2} = 15$) moves to the left (in the direction opposite to $x$-axis direction). We  chose $m_i/m_e=32$ as previous studies \cite{Spitkovsky2008} showed that for  electron-ion mass ratios $m_i/m_e >30$, properties of the shock do not  significantly change with ion mass. Electrons and ions in the incoming plasma beam are assumed to have a negligible energy spread. To reduce the computational effort, the initial contact point of the two counter-propagating streams  is modeled as a reflecting wall at $x=0$~\cite{Spitkovsky2008}. The schematic of the interaction is shown in Fig. \ref{fig0}.  After reflecting from the wall (on the left), the reflected and the incoming plasmas stream through each other and form a collisionless shock. The simulation is performed in the reflecting wall frame, where the downstream (thermalized) plasma behind the shock has a vanishing average flow velocity.
All densities (electron and ion) and fields (electric and magnetic) are expressed in dimensionless units as $\tilde{N}_{i,e} = n_{i,e}/n_0$, $\tilde{B}_y = eB_y/m_e\omega_{pe}c\sqrt{\gamma}$, and $\tilde{E}_{x,z} = eE_{x,z}/m_e\omega_{pe}c\sqrt{\gamma}$. 
%%%%%%%%%%%%%%%%%%%%%%%%%%NEWSECTION%%%%%%%%%%%%%%%%%%%%%%%%%%%%%%%%
\section{MV formation and structure}
The structure of the fully formed shock at $\omega_{pe t}=1140$ is shown in Fig. \ref{fig0}. The transversely averaged density $\tilde{N}(x)=\langle n(x,z)/n_0 \rangle$ (black line; $\langle \rangle$ denotes transverse averaging over the $z$-dimension) and the color plot of the normalized transverse magnetic field ($B_y$) are plotted in Fig.~\ref{fig0}  that was chosen to represent a fully developed shock.
Our focus is on upstream region of the shock. Near the front, the incoming electron and ion flows are cold while the outgoing (reflecting) streams have  considerable  longitudinal and perpendicular momentum spreads (thermal spread).
Due to thermal spread the growth rate of the electron Weibel instability in this region is low as compared to that for cold beam plasmas ($\delta_e\simeq \sqrt{2} \omega_{pe} v_x/c\sim \omega_{pe}$, where $\delta_e$ is the growth rate of electron Weibel instability) 
\cite{Benford,Califano}.
At the same time, the growth rate of the electrostatic two-stream instability  is much smaller than the growth rate of the electron Weibel ($\simeq \omega_{pe}/2\gamma$ in cold limit) \cite{gedalin,gedalin1} since relativistic conditions. Corresponding characteristic wavelength of the electron Weibel instability, $\sim c/\omega_{pe}$,  is $\sim20$
times shorter than that for the two-stream instability.
Electron Weibel instability initiates quickly as the incoming and counter streams meet. The growth rate of electron Weibel instability is found to be $0.13~\omega_{pe}$
that is significantly less than for the cold beam plasma case $\delta_e=\sqrt{2}~\omega_{pe}$ in agreement
with the estimate for a hot electron beam $\simeq\delta_e (1-\Delta \gamma_{\perp}/\gamma)$ \cite{Medvedev1999}, where $\Delta \gamma_{\perp}$
is the transverse energy spread and $\Delta \gamma_{\perp} \backsimeq \gamma$.  Initially, small-scale filaments are formed,  magnetic field grows and then
instability saturates. The maximum value of generated magnetic field is in accordance with estimate for saturation level
$B_{y,s}\simeq \sqrt{\gamma}\,$\cite{Bret2013}.  At the final stage of electron Weibel instability, the electrons in incoming beam are considerably isotropic.

Ion Weibel instability is initiated on background of well thermalized electrons.   The growth rate of ion Weibel instability is found to be close to $0.34~\omega_{pi}$, (here, $\omega_{pi}= \sqrt{4\pi e^2 n_0 /\gamma m_i}$ is the ion plasma frequency)
that is less than in cold approximation  ($\delta_i=\sqrt{2}~\omega_{pi}$)\cite{Stockem2015} due to
the fact that there is considerable initial magnetic field that makes a standard linear analysis not well applicable.
Similar to electrons, ion filaments are forming and magnetic field is growing up to the maximum value $\sim\sqrt{m_i/m_e} B_{y,s}$.
After a rather fast merging, that takes time of order of $\ln(m_i/m_e)\omega_{pi}$ \cite{Stockem2015}, the ion filaments are subject to  such strong nonlinear behavior, as pinching and crossing.
The pinching results in subsequent magnetic field amplification that takes place near the neck of a filament. By this time, a noticeable part of ion kinetic energy is dissipated into the thermal ion energy and supra-thermal particles appear.
Finally, collisionless decay of vortices leads to magnetic field turbalization and chaotisation specially in shock transition region.

The pinching leads to  breakage of an ion current. The characteristic time interval of this process is shrinking with magnetic amplitude growth as it is  inversely proportional to Alfv\'en velocity,  $c_{A}\sim B_y\,$\cite{trubnikov}.
After that, the elongated magnetic vortices of bipolar structure in z-direction are formed similar to those observed for electrons in anisotropic collisionless hydrodynamics \cite{Bychenkov,Yadav2010}.

\begin{figure}
\begin{center}
%\hspace*{-0.4in}
\includegraphics[width=\linewidth]{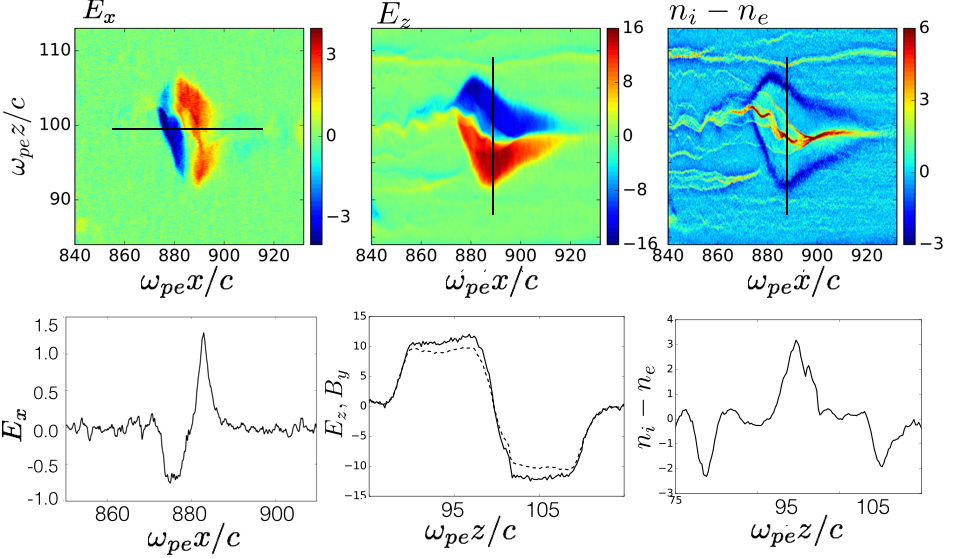}%{field.png}%{fields.png}%{sketch4.png}%{magnetic-d.png}
\end{center}
\caption{a) Distribution of longitudinal electric field ($E_x$) of MV, b) transverse electric field ($E_z$) c) charge density at $\omega_{pe}t=1057$. d) lineout of the longitudinal electric field along line shown in a). e)lineouts of the transverse electric field (dashed line) and out of plane magnetic field (solid line) corresponding to the line shown in b). f) lineout of charge density along the line shown in c)}\label{fig1}
\end{figure}

%%%%%%%%%%%%%%%%%%%%%%%%%
Figure \ref{fig1} shows the structures of longitudinal and transverse electric fields ($E_x$,$E_z$) of nonlinear stage of a typical MV in upstream of the shock at $\omega_{pe}t=1057$, where the MV is developed but not deformed yet. The bottom panel of Fig.\ref{fig1} shows the lineouts of the fields and charge density along black  overlaid lines shown on top panel. Strong electric field is induced around the cavity because of electron evacuation. This field tends to drag the counter stream electrons into the MV. The transverse electric and magnetic  fields ($E_z$ and $B_y$) are  much larger than the longitudinal electric field ($E_x$) as can be seen from Fig. \ref{fig1} . 
The strong Lorentz force $q\textbf{v}\times \textbf{B}$, focuses the incoming ion beam, while expelling the incoming electron beam form the center of MV as  shown in Fig.\ref{fig1}-c. The ion currents are pinched in the self-generated magnetic field. The counter-streaming electron flow follows the ion flow to partly neutralize the beam plasma. However, at the strongly nonlinear stage, significant charge separation appears [see charge density in Fig. \ref{fig1}-c]. The transverse electric field balances the Lorentz force, $E_z\approx v_xB_y/c$. The ion filament pinching results in an increase in the magnetic field and a consequent increase in the Lorentz force, which can be seen in Fig. \ref{fig1}-(b,c) and corresponding lineouts.

%The electron energization mechanism in the upstream region of the shock  is not understood. It is believed that the electric fields around the current filaments slows down the ions and the same field slowing down the ions accelerate the electrons. However, electron energization mechanism needs to be understood. In this letter, we study the electron energization mechanisms in the upstream of the collisionless shock in details. 

\begin{figure}
\begin{center}
%\hspace*{-0.4in}
\includegraphics[width=\linewidth]{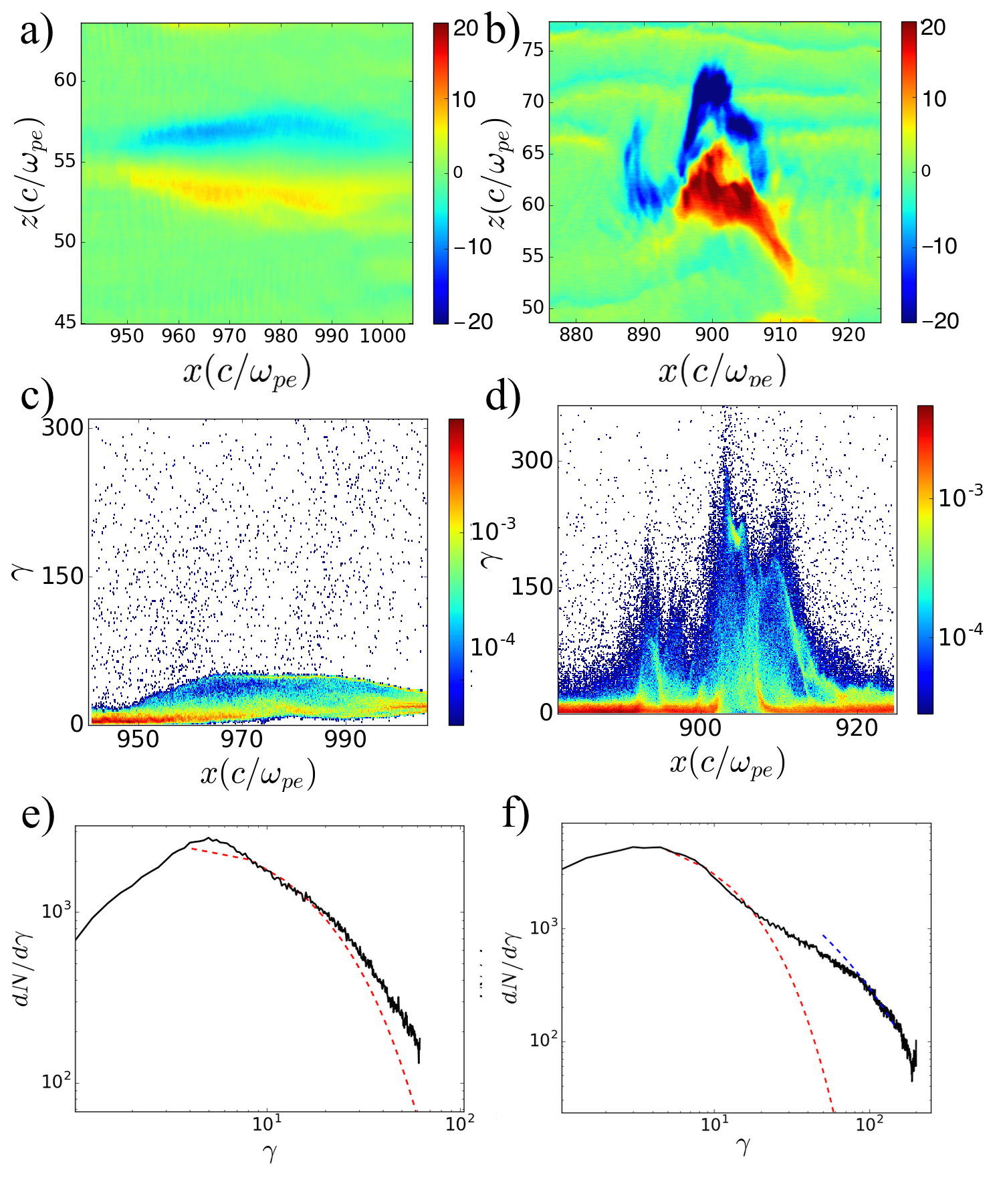}%{sketch4.png}%{magnetic-d.png}
\end{center}
\caption{a,b) Magnetic field distribution evolution of a typical MV at $\omega_{pe}t=1038$ and $1178$. c,d) electron energy distributions averaged over transverse length of MV. e,f) electron energy spectrum evolution corresponding to $\omega_{pe}t=1038$ and $1178$. The low energy part of the electron energy spectrum is fitted by Maxwell–Juttner distribution (red curves) and the non-thermal component at $\omega_{pe}t=1178$ is best fitted by power law (blue). }\label{fig2}
\end{figure}

Figures \ref{fig2}-(a,b) show the distributions of the out of plane plane magnetic fields ($B_y$) at the initial (linear) stage of the MV generation at $\omega_{pe}t=1038$ and  the saturated nonlinear stage of MV at $\omega_{pe}t=1178$  in our simulation. We typically observe  magnetic field enhancement of the MVs  by a factor of $\approx 5 \sim \sqrt{m_i/m_e}$ with respect to the background magnetic field while it propagates towards the shock\cite{naseri2018}. The size of MVs grows to $c/\omega_{pi}$. Figures \ref{fig2}-(c,d) illustrate the electron energy distributions (corresponding to Fig. \ref{fig2}-(a,b), averaged over the transverse size of MD along $z-$ direction. We can see that the electrons gain a large amount of energy at later time. At this time, the transverse electric (and magnetic) field of the MV reaches its maximum. The electron energy has is the largest around the center of the MV and decreases with distance from the center.
%Figure \ref{fig1}-a,b show the distribution of out of plane magnetic field ($B_y$)at $\omega_{pe}t = 1038; 1178$ which shows magnetic dipole evolution. The magnetic field amplitude amplifies more than five times. The distribution of the electron energies are shown in Fig. \ref{fig1}-c,d, which shows the electron spectrum after and before current cut. The electron energy distributions are shown in Fig. \ref{fig1}-e,f. 
The low energy {part of the }electron energy spectrum is fitted by Maxwell–Jüttner distribution with $T_e \approx 10 m_ec^2$ at $\omega_{pe} t = 1038$.
A non-thermal component with energies up to $200 m_e c^2$ appears
at $\omega_{pe}t = 1178$ that is best fitted by a power law with index $\approx 1.5$. The strongest particle acceleration happens during the nonlinear stage of MV formation. The distribution of the electrons is shown in Fig. \ref{fig2}-(e,f).\\
%Figure \ref{fig2a} shows the plasma density (black curve) and counter stream density averaged  in simulation box at $\omega_{pe}t=1038$
% In order to understand the details of the particle acceleration, we tracked the detailed
%motion of the electrons coming in contact with magneticdipole.
 \begin{figure}[h]
\begin{center}
%\hspace*{-0.4in}
\includegraphics[width=\linewidth]{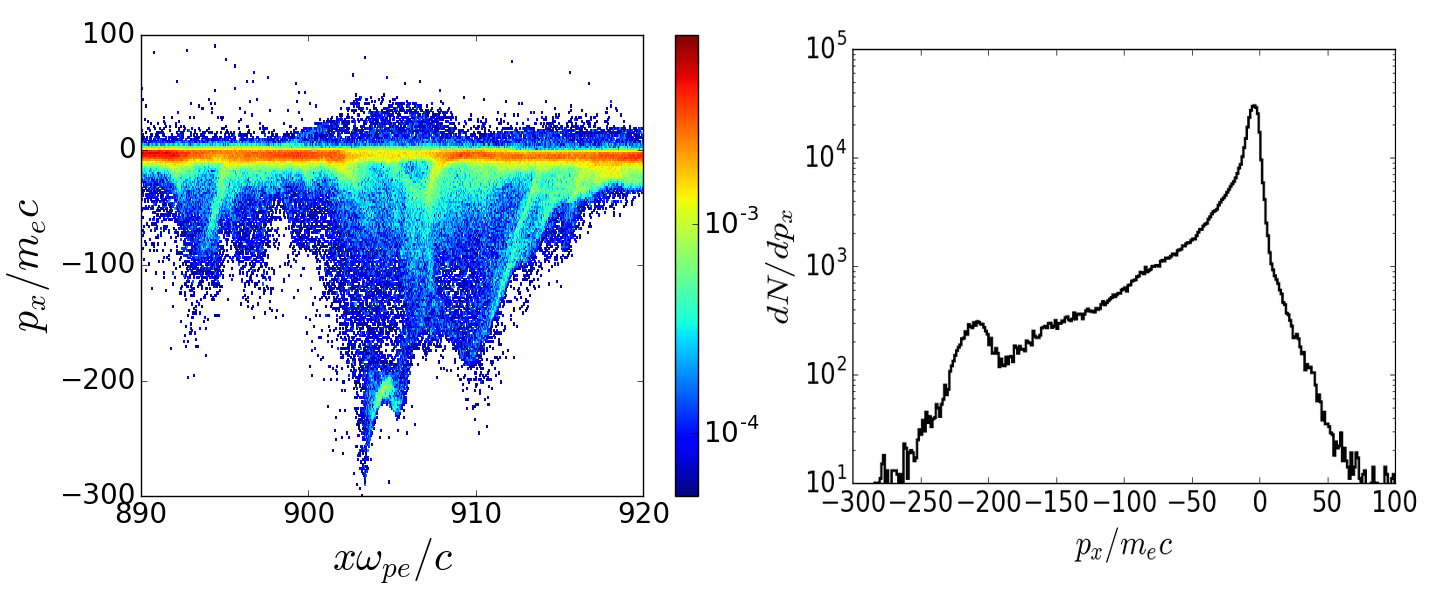}%{field.png}%{fields.png}%{sketch4.png}%{magnetic-d.png}
\end{center}
\caption{a) electron momentum distribution $p_x/m_ec$ along $x$-direction averaged over transverse size of the MV. b) electron phase spectrum corresponding to a) at $\omega_{pe}t=1178$. }\label{fig5}
\end{figure}
\begin{figure}
\begin{center}
%\hspace*{-0.4in}
\includegraphics[width=\linewidth]{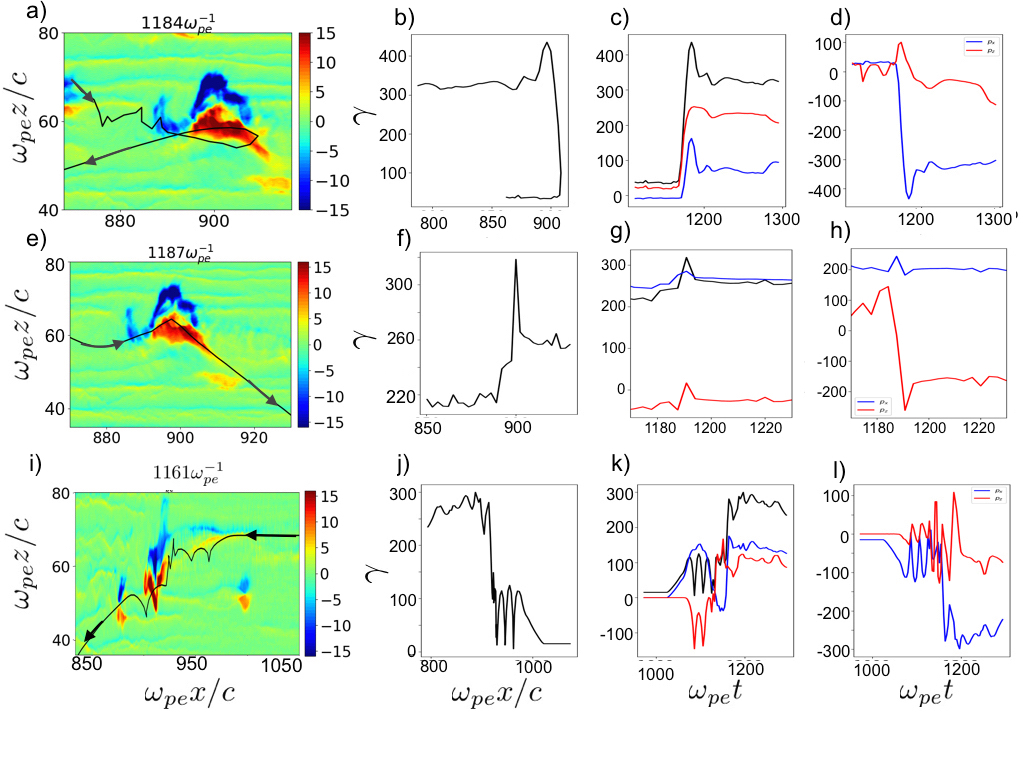}%{prl-traj.jpg}%{field.png}%{fields.png}%{sketch4.png}%{magnetic-d.png}
\end{center}
\caption{Left column shows the distributions of transverse electric field of MV. The trajectories of energized electrons are superimposed and shown by black color.  The middle column shows the electron energy as a function of $x$. The third column illustrates electron energies (black) and work done by longitudinal (blue) and transverse (red) electric fields on the particle as a function of time. The fourth column shows the longitudinal (blue) and transverse (red) momentum ($p_x/mc, p_z/mc$) of electrons as a function of time. }\label{fig4}
\end{figure}\label{fig4}
 %The magnetic field grows during propagation towards downstream. The lobes of the magnetic field expand transversely,
%increasing the vortex radius.
\section{electron energization mechanism}
In the following, we will discuss electron energization mechanisms by magnetic dipole vortices resulting in hot electron tail in electron energy distribution in Fig. 3-e.  that  can also be characterized as  magnetic island, i.e. bipolar magnetic field normal to the current sheet. In order to understand the details of particle acceleration, we tracked the detailed motion of the electrons from the tail of the energy spectrum. The work done by each component of electric field on each particle is calculated throughout the simulation: $W_i=\int_0^t dt' (p_i/\gamma m_{e,i})(\pm e E_i)$, where $i=x,y,z$. Three distinct types of energized electrons were observed in interaction of electrons with MV.  Our analysis of many MVs in the upstream of the shock indicates that $\sim90 \%$ of the energized electron flow from non thermal tail of energy spectrum move towards (return) to the shock transition region after gaining energy from electric fields of MV.  Figure \ref{fig5} shows the electron phase distribution averaged over transverse size of the MV at $\omega_{pe}t=1178$ (See Fig. \ref{fig2}-(d,e). Energetic electrons from the tail of the energy spectrum have negative longitudinal momentum. The longitudinal momentum distribution of the electrons (Fig. \ref{fig5}) averaged over the box shown in Fig. \ref{fig2}-b, illustrates the population of electrons moving towards (or returning) to the shock transition region. The  peak around $p_x\sim -220m_e c$, shown in Fig. \ref{fig5}-b corresponds to the population of electrons shown in Figs. \ref{fig5}-a, \ref{fig2}-d. A few percent of counter stream electrons from the non thermal tail of energy spectrum are pre-accelerated before interaction with MV. These electrons gain some extra energy and continue towards upstream after leaving MV. The third type of energetic electrons are from the incoming electron flow. These electrons trap in linear stage of MV formation and move with MV until the final stage of MV.  Meanwhile they gain energy from the electric fields of MV and  leave MV and move towards shock transition region. \\

We  start with the first type of electron energization mechanism: a typical counter stream electron  moving along $+x$ direction, towards upstream, experiences the magnetic force of $ev_zB_y (-\bf{\hat{x}})$ ($v_z<0, B_y>0$ in this case (Fig.\ref{fig4}-(a,b,c,d) )) along $-x$-direction which is larger than the electric force $-eE_x$ (note that $|E_x| <(|E_z|,|B_y|$, See Fig. \ref{fig1} and supplementary material1). 
%The electron gains energy from longitudinal electric field of the MD as well as transverse electric field.  We see that the longitudinal momentum $p_x$ of the particle changes sign during reflection process. 
 This causes the electron to abruptly turn and  move in the opposite direction towards shock transition region. %which is larger than the electric force $-eE_x$ (note that $|E_x| <(|E_z|,|B_y|$, See Fig. \ref{fig1} and supplementary material1). 
 Meanwhile it gains the energy while moving in the positive lobe of the longitudinal electric field of MV during reflection. The energy gain of electron from longitudinal electric field continues during reflection of the electron ($-eE_xdx>0,(E_x>0, dx<0)$). At the same time the electron gains energy as it moves in transverse electric field ($-eE_z dz, E_z>0,dz<0$ Supplementary material 1), and therefore the energy of the electron increases significantly.
 %This only happens if the transverse electric force on the electron is larger than the magnetic force (this case will be discussed later). 
 As the electron passes the center of the MV and moves towards the negative lobe of longitudinal electric field of the dipole, it loses a fraction of its energy and finally, the electron leaves the MV at its final stage with energy gain of more than an order of magnitude larger than its original energy and moves towards the shock transition region. A typical trajectory of such energetic electron from the tail of the energy spectrum is illustrated in Fig. \ref{fig4}-a overlaid on transverse electric field distribution at $\omega_{pe}t=1132$ showing the return of the electron towards shock transition region. Figure \ref{fig4}-b shows the energy gain of the electron plotted as a function of $x$, showing the energy gain and return of electron  during this process. The evolution of total energy and work done  by the electric field components is plotted in Fig. \ref{fig4}-c. The work done by by both longitudinal and transverse electric field of the MV is leading to the energization of such electrons.  Figure 6 shows energetic electron behavior from tail of the energy spectrum corresponding to energization mechanism discussed here.  Figure 6-a shows the energy gain of such electrons as a function of longitudinal direction ($x$), showing the return of these electrons while gaining a large amount of energy. We can see that electrons interacting with nonlinear stage of MV, where the fields reach their largest magnitude, gain more energy than the electrons interacting with MV at earlier times, when the fields are still growing.  Figure 6-(b,c) show  the energy gain time evolution and  longitudinal momentum of such particles, confirming their return to shock transition region. A characteristic behavior of these energetic electrons is that they return to the shock transition region after interaction with MV. \\
 %This is important because we believe these electrons form the shock transition region in later times and play an important role in stableizing and thinning the shock transition. The electron gains energy from both the longitudinal and transverse electric field of the dipole in its nonlinear stage while the magnetic Lorentz force returns the electron to the shock transition region. { Fig. 6 demonstrates the energy gain for first group of electrons. }\\

%{\bf Neda, I would to characterize what was energy distribution of counter steam electrons before interaction with dipole, as I understand some particles had energy higher that $160 m_ec^2$ before the interaction process.  What is the  temperature of counter stream electrons before interaction with dipole?} 
The second type of energized electrons are the pre-accelerated counter stream electrons moving toward upstream of the shock. These electrons already have large energies ($\gamma_{initial}>160 m_ec^2$ for the typical electron in Fig. \ref{fig4}-e) moving towards upstream prior interacting with nonlinear MV. The transverse magnetic  Lorentz force ($-ev_x B_y(\bf{\hat{z}})$) kicks the electron out of MV. For the typical electron shown in Fig. \ref{fig4}-e $v_x>0, B_y>0$, therefore the magnetic Lorentz force is along $-\bf{\hat{z}}$ and the electron is kicked out of MV (see Supplementary material 2).
The electron loses  energy while moving into MV ($E_z>0,~dz>0$) (Fig. \ref{fig4}-e). Then the magnetic force of MV divert the electron, and the electron gains energy while moving out of MV ($E_z>0, ~dz<0$).  It then continues towards upstream. Figure \ref{fig4}-e shows the trajectory of such electron overlaid on electric field distribution at $\omega_{pe}t=1187$, showing typical electron continue towards upstream after interaction with MV at its nonlinear stage. Figure \ref{fig4}-f  shows the electron energy as a function of $x$ which shows some energy gain for electron before leaving MV. Figure \ref{fig4}-g  shows that most of the energy gain is from transverse electric field.\\ %{\bf Neda, please check graph Figure 5-b (middle panel, third column), according to the graph the most energy comes  from longitudinal field (blue). I suspect the blue should be replaced here by red.  }\\ 
The third type of energetic electrons from the tail of the energy spectrum are the incoming electron flow. The incoming electron enters upstream of the shock and quickly traps in the electric field of MV during linear stage of MV formation. The longitudinal electric field of MV traps the electron, so the electron moves with MV towards shock transition region. At the same time, the electric and magnetic fields of MV grow significantly. The trapped electron reflects from one lobe to the other due to transverse magnetic Lorentz force. Its energy oscillates rapidly between $15$ and $100 m_ec^2$ for typical electron shown in Fig. \ref{fig4}-i (Supplementary material 3). Finally at the final stage of MV, the electron leaves MV while gaining energy mostly from transverse electric field and continues towards shock transition region (Fig. \ref{fig4}-i).\\ We never observed  incoming electrons returning to upstream for obvious reason: the incoming electron flow has mostly longitudinal momentum, therefore the longitudinal force of $-ev_zB_y \bf{\hat{x} }$ is not large enough to return the electron to upstream (because $v_z$ is very small or zero). In addition, the longitudinal electric field force of  $-eE_x \bf{\hat{x}}$ helps electrons to trap in the MV and move with MV. 
\begin{figure}[h]
\begin{center}
%\hspace*{-0.4in}
\includegraphics[width=10cm]{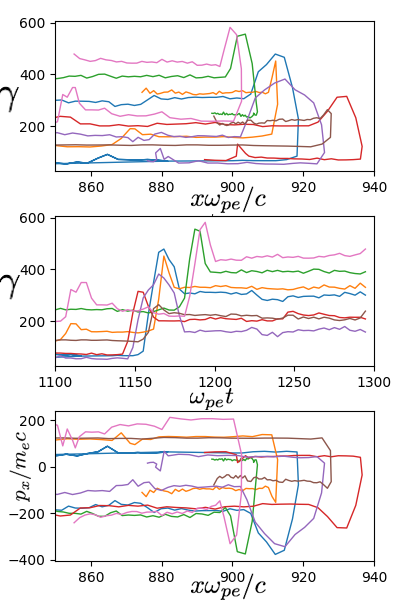}%{field.png}%{fields.png}%{sketch4.png}%{magnetic-d.png}
\end{center}
\caption{a) electron energies ($\gamma$) corresponding to type 1 energization mechanism along longitudinal direction  $x$-direction  b) electron energies as a function of time $\omega_{pe}t$. c) longitudinal momentum of the electrons $p_x/m_ec$ along $x$-direction. }\label{fig6}
\end{figure}

\section{ Conclusions}
Wee studied the process of electron energization in the upstream of electron/ion shock using 2D PIC simulations. Electron energization happens as the electrons (counter-stream and incoming flow) interact with nonlinear stage of MV.  Three distinct processes of electron energization were discussed. This process happens on time scale of dipole evolution and does not require  long times as it is needed for Fermi like acceleration. The fast nonthermal  particles forms power law spectrum. We mention also that this mechanism works for protons.  This investigation will be subject of future publication.    
%\bibliography{sample63}{}
%\bibliographystyle{aasjournal}
%\begin{references}
  %\reference {Piran} T. Piran, Rev. Mod. Phys. \textbf{76} (2004).

%\end{references}

%\bibliography{sample63}{}
%\bibliographystyle{aasjournal}



\end{document}